\documentclass[aps,twocolumn,showpacs]{revtex4}
\usepackage{graphics}
\usepackage{graphicx}

\begin{document}
\title{Transient response under ultrafast interband excitation of an
intrinsic graphene }
\author{P.N. Romanets}
\author{F.T. Vasko}
\email{ftvasko@yahoo.com}
\affiliation{Institute of Semiconductor Physics, NAS of Ukraine,
Pr. Nauky 41, Kiev, 03028, Ukraine}
\date{\today}

\begin{abstract}
The transient evolution of carriers in an intrinsic graphene under ultrafast
excitation, which is caused by the collisionless interband transitions, is
studied theoretically. The energy relaxation due to the quasielastic acoustic
phonon scattering and the interband generation-recombination transitions due
to thermal radiation are analyzed. The distributions of carriers are obtained
for the limiting cases when carrier-carrier scattering is negligible and when
the intercarrier scattering imposes the quasiequilibrium distribution. The
transient optical response (differential reflectivity and transmissivity) on a
probe radiation and transient photoconductivity (response on a weak dc field)
appears to be strongly dependent on the relaxation and recombination dynamics
of carriers.
\end{abstract}


\maketitle
\section{Introduction}
The transient response of photoexcited carriers under ultrafast
interband pumping has been studied during the last decades in bulk
semiconductors and heterostructures (see Ref. 1 for review). The
unusual transport of carriers in graphene is caused by a
neutrinolike energy spectrum in gapless semiconductor, which is
described by the Weyl-Wallace model \cite{2}, and a substantial
modification of scattering processes. Recently, the properties of
graphene after ultrafast interband excitation attract special
attention. The experimental results in relaxation dynamics of
photoexcited electrons and holes were published in \cite{3,4,5} and
\cite{6} for epitaxial and exfoliated graphene, respectively. The
relaxation of nonequilibrium optical phonons, which are emitted by
carriers after photoexcitation, is studied in Ref. 7. The
theoretical consideration of the carrier relaxation and
generation-recombination processes caused by optical phonons is performed
in \cite{8,9}. The quasielastic energy relaxation of carriers due to
acoustic phonons was considered in \cite{10,11} for low energy
carriers (at low temperatures or under mid-IR excitation). In
particular, an interplay between energy relaxation and
generation-recombination processes determines the relaxation
dynamics of photoexcited carrier distribution. \cite{10} To the best
of our knowledge, both this interplay and the relaxation dynamics at
low temperatures are not considered so far. Thus, the investigation
of the transient response of carriers under these conditions is
timely now.

In this paper, we consider the transient response of an intrinsic
graphene in case of ultrafast interband excitation in passive
region, where the carrier energies are smaller than the optical phonon
energy. Such a regime can be realized under the pumping in the
mid-infrared (IR) spectral region or at low temperatures, when the
peak of photoexcited carriers formed after the process of optical
phonon emission, remains a narrow one. Describing the
photoexcitation process, we restrict ourselves by the collisionless
regime, when a pulse duration, $\tau_p$, is shorter than the
momentum relaxation time. Considering the low-temperature transient
dynamics of photoexcited carriers, one takes into account the
intraband quasielastic energy relaxation due to acoustic phonons
and generation-recombination interband transitions due to thermal
radiation. The carrier-carrier scattering is described
within two limiting regimes: ($i$) when the Coulomb interaction is
unessential, and ($ii$) when intercarrier scattering imposes the
quasiequilibrium distribution of carriers. With the obtained
transient distribution of carriers, we analyze a time-dependent
response on the probe field, i.e. we consider the transient
reflection and transmission in THz and mid-IR spectral regions. The
transient photoconductivity is also analyzed below, because the
energy relaxation corresponds to a nanosecond scale (the radiative
recombination remains essential up to microsecond).

Since the electron-hole energy spectrum and scattering processes are
symmetric in an intrinsic graphene, the phenomena under
consideration are described by the same distribution functions for
electrons and holes, $f_{pt}$. Such distribution is governed by the
general kinetic equation \cite{12}:
%1
\begin{equation}
\frac{\partial f_{pt}}{\partial t} =\sum_k J_k\left\{ f_t|p\right\}
+G\{ f|pt\} ,
\end{equation}
where the collision integrals $J_k\left\{ f_t|p\right\}$ describe
the relaxation of carriers caused by the carrier-carrier scattering
($k=cc$), the acoustic phonons ($k=ac$), and the thermal radiation
($k=r$), respectively. The photogeneration rate, $G\{ f|pt\}$,
describes the interband excitation of electron-hole pairs by the
mid-IR ultrafast pulse. Below Eq. (1) is solved with the initial
condition $f_{pt\to -\infty}=f_p^{(eq)}$, where $f_p^{(eq)}$ is the
equilibrium distribution. The transient response on a probe
radiation is described by the dynamic conductivity due to interband
transitions. The transient response on a weak dc field (photoconductivity)
is considered with the use of the phenomenological model of momentum
scattering suggested in \cite{13}.

The analysis carried out  below is organized as follows. The
photoexcitation process under the inerband pumping is described in
Sec. II. The transient evolution distributions are
given in Sec. III for the cases ($i$) and ($ii$). Section IV
presents a set of results of transient reflectivity and
transmittivity, and also the transient photoconductivity. The
discussion of the assumptions used and concluding remarks are given
in the last section. Appendix contains the microscopical
evaluation of the interband photogeneration rate under ultrafast
interband excitation.

\section{Ultrafast excitation}
In the framework of the Weyl-Wallace model (spin- and
valley-degenerate linear energy spectrum of carriers which is
determined by the characteristic velocity $v_W$), the interband
photoexcitation is caused by the in-plane electric field,
$w_t{\bf E}\exp (-i\Omega t)+c.c.$ where $\bf E$ is the field strength,
$\omega$ is the frequency, and $w_t$ is the envelope form-factor.
Eq. (1) is transformed to the collisionless form on the initial
intervals, when scattering mechanisms are not essential: $\partial
f_{pt} /\partial t = G\left\{ {f|pt} \right\}$. Using the boundary
condition of Eq. (1), one can rewrite this equation in the integral
form $f_{pt}=f_p^{(eq)}+\int_{-\infty}^t dt'G\{ f|pt'\}$. The
photogeneration rate here is evaluated in Appendix as follows:
%2
\begin{eqnarray}
G\left\{ f|pt\right\}=\left( \frac{eEv_W}{\hbar\Omega}\right)^2 w_t
\int\limits_{-\infty}^0 d\tau w_{t+\tau} \nonumber \\
\times\cos\left[\left(\frac{2v_W p}{\hbar}-\Omega\right)\tau\right]
\left( 1-2f_{pt+\tau}\right) ,
\end{eqnarray}
where the Pauli blocking factor $(1-2f_{pt+\tau})$ is responsible for the
coherent Rabi oscillations of the excited carriers.

Introducing the dimensionless intensity, $I_{ex}=(eE\tau_pv_W
/\hbar\Omega )^2$, we consider below the linear regime of excitation
which takes place if $I_{ex}\ll 1$ and $f_{pt}\ll 1$, so that the Pauli
factor can be neglected (if $\hbar\Omega$ comparable to the
equilibrium temperature $T$ one has to use the equilibrium Pauli
factor in Eq. (2)). Using the Gaussian form-factor
$w_t=\sqrt[4]{2/\pi}\exp\left[ -\left( t/\tau_p\right)^2\right]$
with the pulse duration $\tau_p$, \cite{14} one obtaines the
photoexcited distribution in the form
%3
\begin{equation}
f_{p t}^{(ex)}\approx I_{ex}\int\limits_{-\infty}^{t}{dt'} w_{t'}
\int\limits_{-\infty}^0 {d\tau w_{t'+\tau}\cos\left( {\frac{{2v_W \delta p}}
{\hbar }\tau } \right)} .
\end{equation}
Here $\delta p=p-p_\Omega$ is centred in the characteristic momentum
$p_\Omega =\hbar\Omega /2v_W$. The evolution of photoexcited
distribution, $f_{p t}^{(ex)}/I_{ex}$, is shown in Fig. 1. The
distribution is dependent on $t/\tau_p$ and $\delta p/\Delta p$,
where $\Delta p=\hbar /2v_W\tau_p$ determines the width of
distribution which is proportional to $\tau_p^{-1}$. For
$t\gg\tau_p$, the integrations in Eq. (3) can be exactly performed
and we obtain steady-state distribution after the photoexcitation
pulse, $f_{p}^{(ex)}= f_{p t\to\infty}^{(ex)}$, as the Gaussian peak
of width $\propto\Delta p$:
%4
\begin{equation}
f_{p}^{(ex)}=\sqrt{\frac{\pi}{2}}I_{ex}e^{-(\delta p/\sqrt{2}\Delta p)^2} .
\end{equation}
Thus, at $t\geq 2\tau_p$ (see Fig. 1) one can omit the
photogeneration rate in Eq. (1) using instead the initial condition:
%5
\begin{equation}
f_{pt=0}=f_{p}^{(eq)}+f_{p}^{(ex)} ,
\end{equation}
which is given as a sum of the equilibrium and photoexcited
contributions. The condition (5) can be used directly in case of
weak intercarrier scattering. In case of optical excitation, with a
subsequent emission of cascade of $2{\cal N}$ optical phonons of
energy $\hbar\omega_0$, the photoexcited distribution can be
written in the form (5) where $\delta p$ is centred in
$p_{\overline{\omega}}= (\hbar\Omega -2{\cal N}\hbar\omega_0)/2v_W$
and $\Delta p$ is included an additional broadening during the
cascade emission.

\begin{figure}[ht]
\begin{center}
\includegraphics{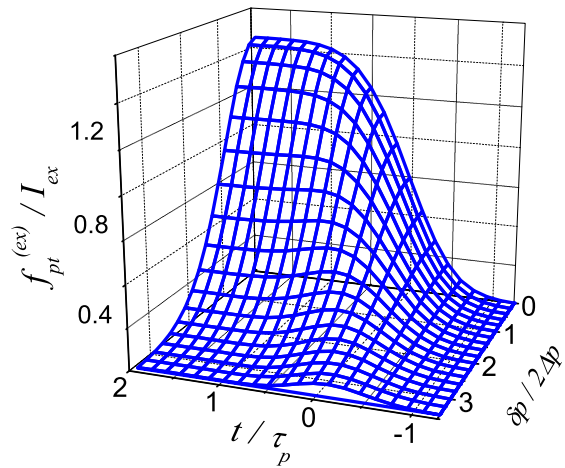}
\end{center}\addvspace{-1 cm}
\caption{Temporal evolution of photoexcited distribution $f_{pt}^{(ex)}$
normalized to $I_{ex}$ versus dimensionless momentum and time,
$\delta p/2\Delta p$ and $t/\tau_p$.}
\end{figure}

Under an effective intercarrier scattering, one needs to calculate
the initial temperature and concentration of carriers. The photoexcited
concentration and energy of carriers, which are described by the peak of
distribution (4) are given by
%6
\begin{equation}
\left|\begin{array}{*{20}c} \Delta n_{ex} \\  \Delta E_{ex}\end{array}\right| =
\frac{4}{L^2}\sum\limits_{\bf p}\left|\begin{array}{*{20}c} 1 \\ {v_W p}
\end{array} \right|f_p^{(ex)} \simeq\frac{I_{ex} (\overline{\omega}/v_W )^2}
{2\overline{\omega}\tau _p}\left|\begin{array}{*{20}c} 1 \\ \hbar\overline{\omega}/2
\end{array}\right| ,
\end{equation}
where $L^2$ is the normalization area.  One obtains $\Delta
E_{ex}/n_{ex}=\hbar\overline{\omega}/2$, for the Gaussian shape of
pulse, i.e. the averaged energy per generated particle is equal to
the excitation energy. In case of optical excitation, with $\cal N$
optical phonons emitted, the energy per photoexcited particle,
$\Delta E_{ex}/\Delta n_{ex}$, agrees closely with $\hbar\omega
-{\cal N}\hbar\omega_0$ (see above).

\begin{figure}[ht]
\begin{center}
\includegraphics{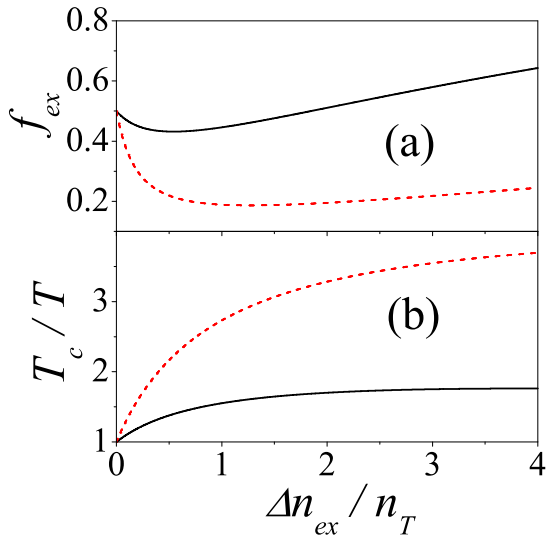}
\end{center}\addvspace{-1 cm}
\caption{Initial  maximum distribution (a) and effective temperature
(b), $f_{ex}$ and $T_{ex}$, versus pumping ($\Delta
n_{ex}/n_T\propto I_{ex}$ for $\hbar\overline{\omega}=60$ meV and
120 meV (solid and dashed curves, respectively). }
\end{figure}

If $\tau_p\ll\tau_{cc}\ll\tau_{ac,r}$ ,
where $\tau_{cc}$, $\tau_{ac}$, and $\tau_{r}$ correspond to the
intercarrier scattering, the energy relaxation, and the
generation-recombination processes, respectively [the
Coulomb-controlled case ($ii$)], the dominanting carrier-carrier
scattering imposes the quasi-equilibrium distribution
%7
\begin{equation}
f_{pt}=\left[\exp\left( \frac{v_W p-\mu_t}{T_t}\right) +1\right]^{-1}
\end{equation}
with the effective temperature $T_t$ and the quasichemical potential
$\mu_t$. If $\tau_{cc}\ll t\ll\tau_{ac,r}$, the initial values
$T_{ex}=T_{t\to 0}$ and $f_{ex}=f_{p=0t\to 0}$ are determined from
the concentration and energy conservation requirements:
%8
\begin{equation}
\frac{2}{\pi}\left(\frac{T_{ex}}{\hbar v_W}\right)^2 \int\limits_0^\infty
dxxf_{x} \left| {\begin{array}{*{20}c}  1  \\  {T_{ex} x}  \\
\end{array}} \right| = \left| {\begin{array}{*{20}c}
   {n_T  + \Delta n_{ex} }  \\  {E_T  + \Delta E_{ex} }  \\
\end{array}} \right| ,
\end{equation}
where the function $f_{x}$ is introduced according to $f_{x}\equiv
f_{ex} /[e^x(1-f_{ex})+f_{ex}]$. Using $\Delta n_{ex}$ and $\Delta
E_{ex}$ given by Eq. (6) and solving the transcendental system (8)
one obtains the initial values $f_{ex}$ and $T_{ex}$.  The
calculations here and below are performed for the nitrogen
temperature, $T=$77 K, the excitation energies
$2v_Wp_{\overline{\omega}}=$120 meV (CO$_2$ laser) and 60 meV (as an
example of interband excitation with subsequent optical phonon emission),
and the broadening energy $\hbar /\tau_p\simeq$6.6 meV, which
 corresponds to the pulse duration $\simeq$0.1 ps. In Fig. 2 we plot
$f_{ex}$ and $T_{ex}$ versus the pumping level which is proportional
to $\Delta n_{ex}/n_T$.  Fast increase of $T_{ex}$ and fast decrease
of $f_{ex}$ take place for $\Delta n_{ex}/n_T< 1$, while a linear
increase of these values are realized if $\Delta n_{ex}/n_T>1$.

\section{Energy relaxation and recombination}
In this section we analyze the transient evolution of $f_{pt}$
caused by the energy relaxation and recombination processes. We
consider the cases ($i$) and ($ii$), when the initial condition is
given by Eq. (5) and written through $f_{ex}$ and $T_{ex}$ plotted
in Fig. 2.

\subsection{Weak intercarrier scattering}
If the carrier-carrier scattering is ineffective [case ($i$)], the
distribution $f_{pt}$ is governed by the kinetic equation (1)
without the $cc$-contribution
%9
\begin{eqnarray}
\frac{\partial f_{pt}}{\partial t}=\frac{\nu_p^{\ss (qe)}}{p^2}\frac{d}{dp}
\left\{ p^4\left[\frac{df_{pt}}{dp}+\frac{f_{pt}(1 - f_{pt})}{p_T}\right]
\right\} \nonumber \\
+\nu_p^{(r)}[N_{2p/p_T}(1 - 2f_{pt})-f_{pt}^2]
\end{eqnarray}
and with the initial condition (5) used instead of generation rate.
Here we substituted the explicit expressions of the collision
integrals for the quasielastic acoustic scattering approximation
(written in the Fokker-Planck form) and for the
generation-recombination processes, see discussion in \cite{10}. The
Planck distribution $N_{2p/p_T}$ is written through $p_T=T/v_W$
while the energy relaxation rate $\nu_p^{(qe)}=v_{qe}p /\hbar$ and
the rate of radiative transitions $\nu_p^{(r)}=v_rp /\hbar$ are
written through the characteristic velocities $v_{qe}\propto T$ and
$v_r$ \cite{15}.

The boundary conditions are imposed by both the condition $f_{p\to\infty t}=0$,
which is transformed into the requirement
%10
\begin{equation}
p^4\left(\frac{\partial f_{pt}}{\partial p}+\frac{f_{pt}}{p_T}
\right)_{p\to\infty}< {\rm const} ,
\end{equation}
and Eq. (9) at $p=0$ which is transformed into the initial condition
$f_{p=0t}=1/2+f_{p=0}^{(ex)} \exp [-(v_r/v_W)Tt/\hbar ]$. According
to Eq. (4) one obtains $f_{p=0}^{(ex)}=\sqrt{\pi /2}I_{ex}\exp
[-(\Omega\tau_p)^2/2]\ll 1$ and one can neglect the second
contribution in this initial condition, so that $f_{p=0t}=1/2$.
Numerical solution of the Cauchy problem given by Eqs. (5), (9), and
(10) is obtained below by the use of the iteration procedure. \cite{16}
\begin{figure}[ht]
\begin{center}
\includegraphics{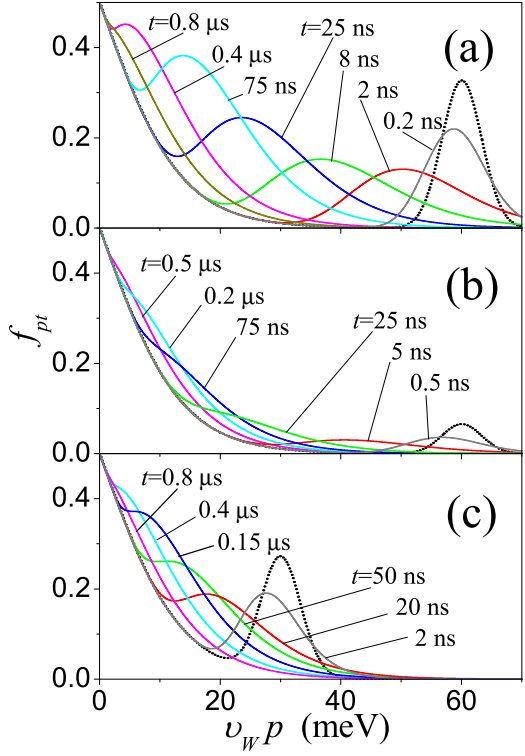}
\end{center}\addvspace{-1 cm}
\caption{Distribution $f_{pt}$ versus carrier energy $pv_W$ for
different delay times (marked) and excitation conditions: (a)
$\hbar\overline{\omega}=$120 meV and $I_{ex}=$0.26, (b)
$\hbar\overline{\omega}=$120 meV and $I_{ex}=$0.05, and (c)
$\hbar\overline{\omega}=$60 meV and $I_{ex}=$0.21. }
\end{figure}

In Fig.3 we demonstrate the evolution of the distribution $f_{pt}$
at 77 K for the cases when carriers are excited around the energies
60 meV and 30 meV. The delay times are marked in panels a-c and the
pumping levels are determined through the
initial peak value, given by $\sqrt{\pi /2}I_{ex}$, see Eq. (4)).
Under mid-IR pumping with pulse duration $\tau_p=$0.1 ps and the
spot sizes $\sim$0.5 mm the above-used pumping levels correspond to
the pulse energies $\sim$85 pJ and $\sim$17 pJ for Figs. 3a and 3b,
respectively, see \cite{17} for experimental details. Under optical
pumping ($\hbar\Omega\sim$1.6 eV) and subsequent emission of phonon
cascade, the pumping level in Fig. 3c corresponds to the pulse
energy $\sim$12 nJ (duration and size are the same as above). One can
see that the transient evolution of distribution occurs in two
stages: energy relaxation and recombination. During the first stage
(about $t\lesssim$50 ns, which is dependent on position and maximum
value $f_p^{(ex)}$; compare with Figs. 3a-c) the initial peak is
tranformed into the quasiequilibrium high-energy tail (with the
equilibrium temperature caused by the energy relaxation) which is
connected to the low-energy equilibrium distrbution. During the next
stage (up to 1 $\mu$s) the high-energy tail shifts to the lower
energies and transforms into the equilibrium distribution due to
 effective radiative recombination in low-energy region.

\subsection{Coulomb-controlled case}
In the carrier-carrier scattering case ($ii$), one has to describe
the transient evolution of the effective temperature $T_t$ and the
maximum distribution $f_t=f_{p=0t}$, that replaces the chemical potential.
Since the intercarrier scattering change neither the concentration,
$n_t=(4/L^2)\sum_{\bf p}f_{pt}$, nor the energy of carriers,
$E_t=(4/L^2)\sum_{\bf p}v_Wpf_{pt}$, the balance equations for $n_t$
and $E_t$ take forms: \cite{18}
%11
\begin{equation}
\frac{d}{{dt}}\left| {\begin{array}{*{20}c} {n_t } \\ {E_t }  \\
\end{array}} \right| = \frac{4}{{L^2 }}\sum\limits_{\bf p} {\left|
{\begin{array}{*{20}c} {J_r \{ f_t |p\} }  \\
   {v_W p\left[ {J_{ac} \{ f_t |p\}  + J_r \{ f_t |p\} } \right]}  \\
\end{array}} \right|} .
\end{equation}
Further, we transform the balance equations, expressing the
left-hand side of (11) through $T_t$ and $f_t$ as follows:
%12
\begin{eqnarray}
\frac{d}{dt}\left( T_t^2A_t^{(1)}\right) =R_t^{(1)} , \\
\frac{d}{dt}\left( T_t^3A_t^{(2)}\right)=R_t^{(2)}+Q_t .
\nonumber
\end{eqnarray}
Here the coefficients $A_t^{(1,2)}$ are written as
$A_t^{(q)}=\int_0^\infty dxx^qf_{xt}$, where the quasiequilibrium
distribution is given by $f_{xt}=f_t/\left[ e^x (1-f_t)+f_t\right]$,
so that $A_t^{(q)}/T_t^l$ are only depend on $f_t$. After
substitution of the collision integrals $J_r$ \cite{10, 18} and
integration, the generation-recombination contributions to Eq. (12)
are obtained in the form
%13
\begin{equation}
R_t^{(q)}=\frac{2v_rT_t^{q+2}}{v_W\hbar}\int\limits_0^\infty dxx^{q+2}f_{xt}^2
\left[\frac{e^{2x}(1-f_t)^2}{(e^{x2T_t/T}-1)f_t^2}-1\right] .
\end{equation}
Similarly, the energy relaxation contribution is written by the use
of $J_{ac}$ as follows
%14
\begin{equation}
Q_t=\frac{T-T_t}{T}\frac{v_{qe}T_t^4}{v_W\hbar}\int\limits_0^\infty dxx^4
e^x f_{xt}^2\frac{1-f_t}{f_t} .
\end{equation}
The initial conditions for the system (11) are written as
$T_{t=0}=T_{ex}$ and $f_{t=0}=f_{ex}$.

\begin{figure}[ht]
\begin{center}
\includegraphics{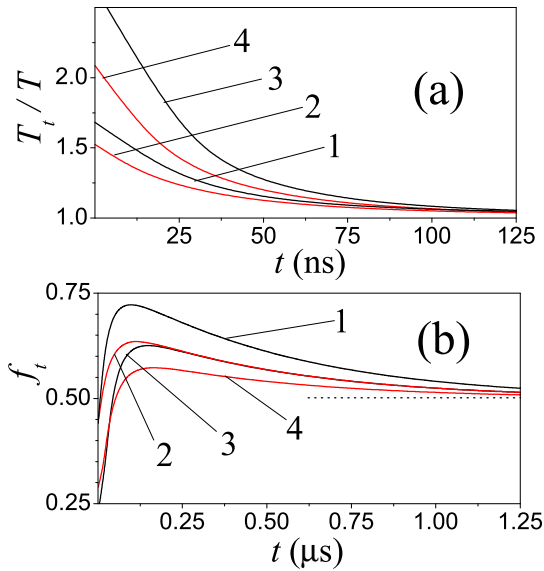}
\end{center}\addvspace{-1 cm}
\caption{Temporal evolution of effective temperature, $T_t$ (a), and
maximum distribution, $f_t$ (b), for  different excitation
conditions: (1) $\hbar\overline{\omega}=$60 meV and $I_{ex}=$0.21,
(2) $\hbar\overline{\omega}=$60 meV and $I_{ex}=$0.1, (3)
$\hbar\overline{\omega}=$120 meV and $I_{ex}=$0.052, and (4)
$\hbar\overline{\omega}=$120 meV and $I_{ex}=$0.026.}
\end{figure}

\begin{figure}[ht]
\begin{center}
\includegraphics{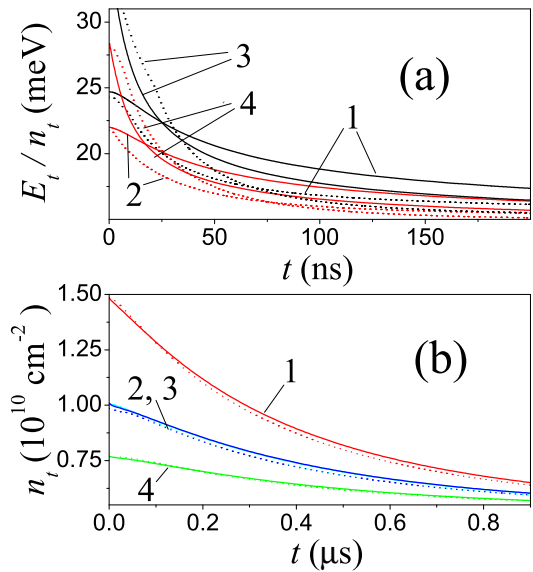}
\end{center}\addvspace{-1 cm}
\caption{Energy per carrier (a) and concentration (b) versus time.
Solid and dotted curves  correspond to the cases  ($i$) and ($ii$) ,
respectively; excitation conditions (1) - (4) are the same as in
Fig. 4.}
\end{figure}

Numerical solution of the nonlinear system (11) is performed using
the iteration procedure. In Fig. 4 we plot $T_t/T$ and $f_t$ versus
time. Temperature relaxes to the equilibrium one during the energy
relaxation times ($\lesssim$ 100 ns) while $f_t$, which is
determined by the chemical potential $\mu_t$, relaxes to 1/2 over
1 $\mu$s (the recombination time scale), in analogy with the
case ($i$). Notice, that after the fast energy relaxation, one
obtains $f_t>$1/2 [dotted line in Fig. (4b)], i.e. the low-energy
electron-hole pairs appear to be unstable. \cite{19}

Fig. 5 shows the plot of temporal evolutions of the energy per particle
and concentration, $E_t/n_t$ and $n_t$ [see the definitions before Eq.
(11)], for the cases ($i$) and ($ii$). The relaxation processes to
the equilibrium (at nitrogen temperature, $E_{t\to\infty}/
n_{t\to\infty}\simeq$14.5 meV and $n_{t\to\infty}\simeq5.3\cdot 10^9$
cm$^{-2}$) occur during the same scales as in Figs. 3 and 4. The
temporal dependencies of $n_t$ obtained for both cases are in good
agreement (the carreir-carrier scattering does not change
concentration) while $E_t$ demonstrates a different evolution for
cases ($i$) and ($ii$) at $t<$50 ns. This is because of drift and
decrease of photoexcited peak during the energy relaxation time, see
Fig. 3.

\section{Transient response}
Here we turn to consideration of the response of photoexcited
carriers on a probe radiation (reflection and transmission in the
THz and mid-IR spectral regions) and on a weak dc electric field
(photoconductivity). The transient electrodynamics of graphene is
described using the time-dependent dynamic conductivity,
$\sigma_{\omega t}$, which is caused by the collisionless interband
transitions, see Appendix B. The transient photoconductivity is
calculated by the use of the phenomenological model of momentum
relaxation suggested in \cite{13}.

\subsection{Reflection and transmission}
To calculate the transient reflectance and transmittance of the
graphene sheet placed at $z=0$ on the in-plane electric field ${\bf
E}_{zt}\exp (-i\omega t)$ propagated along $0Z$, we apply the wave
equation, see \cite{20} and references therein. The induced current
density, $\sigma_{\omega t}E_{z=0}$, is located around $z=0$ and
 direction of in-plane field ${\bf E}_{zt}$ is not essential due to
the in-plane isotropy of the problem. Separating the incident
radiation, $E_{in}e^{ik_\omega z}$, with the wave vector $k_\omega
=\omega /c$, we write the field distribution outside of the graphene
sheet in the form:
%15
\begin{equation}
E_{zt} =\left\{ \begin{array}{*{20}c} E_{in}e^{ik_\omega z}+E_t^{(t)}
e^{-ik_\omega z} , & {z<-0}  \\ E_t^{(t)} e^{i\overline{k}_\omega z} ,
& {z>+0}  \end{array} \right. ,
\end{equation}
where $\overline{k}_\omega =\sqrt{\epsilon}\omega /c$ is the wave vector
in the substrate with the dielectric permittivity $\epsilon$. The
transmitted and reflected electric fields, $E_t^{(t)}$ and $E_t^{(r)}$,
are determined from the boundary conditions at $z\to 0$ as follows:
%16
\begin{equation}
\frac{E_t^{(t)}}{E_{in}}=\frac{2}{1+A_{\omega t}}, ~~~~
\frac{E_t^{(t)}}{E_{in}}=\frac{1 - A_{\omega t}}{1+A_{\omega t}} .
\end{equation}
Here we introduce the dimensionless factor $A_{\omega t}=\sqrt\epsilon +
(4\pi /c)\sigma_{\omega t}$. The reflection and transmission coefficients,
$R_{\omega t}=|E_t^{(r)}|^2/E_{in}^2$ and $T_{\omega t}=|E_t^{(t)}|^2/E_{in}^2$,
are written through $A_{\omega t}$ according to
%17
\begin{equation}
R_{\omega t}=\left|\frac{1-A_{\omega t}}{1+A_{\omega t}}\right|^2 , ~~~~~
T_{\omega t}=\frac{4\sqrt \epsilon}{\left| 1+A_{\omega t}\right|^2} .
\end{equation}
Using $\sigma_{\omega t}$ determined by Eqs. (B3) and (B4), we consider below
the differential changes in reflectivity and transmissivity,
$(\Delta R/R)_{\omega t}=(R_{\omega t}-R_\omega^{(eq)})/R_\omega^{(eq)}$ and
$(\Delta T/T)_{\omega t}=(T_{\omega t}-T_\omega^{(eq)})/T_\omega^{(eq)}$,
which are written through the equilibrium reflection and transmission
coefficients, $R_\omega^{(eq)}$ and $T_\omega^{(eq)}$.

\begin{figure}[ht]
\begin{center}
\includegraphics{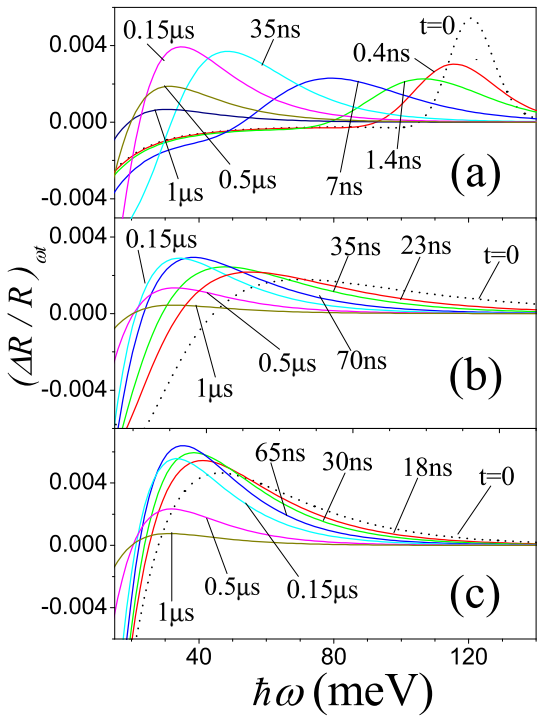}
\end{center}\addvspace{-1 cm}
\caption{(a) Spectral dependencies of differential reflectivity,
$(\Delta R/R)_{\omega t}$, for different delays (marked) at the
excitation conditions: (a) $\hbar\overline{\omega}=$120 meV and
$I_{ex}=0.052$ in the case ($i$), (b) $\hbar\overline{\omega}=$120 meV
and $I_{ex}=$0.052 in the case ($ii$), and (c) $\hbar\overline{\omega}=$60
meV and $I_{ex}=$0.21 in the case ($ii$). }
\end{figure}

The evolution of the differential reflectivity for the cases ($i$)
and ($ii$) are shown in Figs. 6a and 6b, 6c, respectively. If the
Coulomb scattering is not effective [case ($i$)], the distribution
of carriers relaxes during the energy relaxation time scale (around
10 ns, cf. with Fig. 3), when a quenching of photoexcited peak takes
place (if $\hbar\omega$ is comparable with the peak energy).
In case ($ii$) any peculiarities of the spectrsal dependencies at
stort times are absent because the initial distribution is
transformed into the quasiequilibrium one during times
$\sim\tau_{cc}\to 0$. The further evolution of $(\Delta R/R)_{\omega
t}$ is limited by the generation-recombination process and extended
up to microseconds. In the THz spectral region ($\hbar\omega\geq$10
meV is considered here because we neglect the intraband relaxation),
the differential reflectivity increases and changes a sign. In the
high-energy region, $(\Delta R/R)_{\omega t}$ decreases monotonically
with $\omega$ and $t$ and does not exceed $\sim 10^{-4}$ for the
near-IR spectral region. Beside of this, the response is approximately
proportional to the pumping intensity, $I_{ex}$, and $(\Delta R/R)_{\omega t}$
increases with increasing of the photoexcitation
energy, $\hbar\overline{\omega}$ (cf. Figs. 6b and 6c).

\begin{figure}[ht]
\begin{center}
\includegraphics{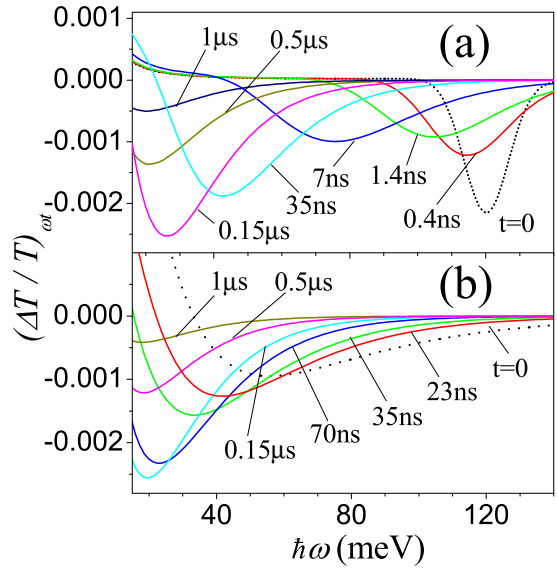}
\end{center}
\addvspace{-1 cm} \caption{(a) Differential transmissivity, $(\Delta
T/T)_{\omega t}$, versus $\hbar\omega$ and $t$ for cases ($i$) and
($ii$) [panels (a) and (b), respectively] at the same excitation
conditions: $\hbar\overline{\omega}=$120 meV, $I_{ex}=$0.052. }
\end{figure}
In Fig. 7 we plot the differential transmissivity for the cases
($i$) and ($ii$) under the same excitation conditions.
Once again, in the high-frequency region the
differential transmissivity decreases slowly (during a microsecond
time scale) and $(\Delta T/T)_{\omega t}$ does not exceed
$\sim 10^{-4}$ for the near-IR spectral region. In the THz
spectral region, $(\Delta T/T)_{\omega t}$ increses and changes the
 sing in the same manner as $(\Delta R/R)_{\omega t}$ (cf. Figs. 6
and 7). The dependencies on the excitation parameters ($I_{ex}$ and
$\hbar\overline{\omega}$) are also similar to the reflectivity.
Additionally, in case ($i$) a fast (at $t<$10 ns) quenching of the
photoexcited peak contribution in the spectral region
$\sim\hbar\Omega$ takes place.

\subsection{Photoconductivity}
Finally, we consider the transient photoconductivity, i.e. the
response of the photoexcited carriers to the weak dc electric field.
Since the momentum relaxation is governed by elastic scattering
mechanisms, \cite{13} one can use the following expression for the
dc conductivity~$\sigma_t$:
%18
\begin{equation}
\sigma_t =\sigma_0\left[ 2f_{p=0t}-\frac{l_c}{\hbar}\int_0^{\infty}dpf_{pt}
\frac{\Psi '(pl_c/\hbar)}{\Psi (pl_c/\hbar)^2}\right] .
\end{equation}
Here $l_c$ is the correlation length characterizing the disorder
scattering and the function $\Psi (z)=e^{-z^2}I_1(z^2)/z^2$ is
written through the first order Bessel function of imaginary
argument, $I_1(z)$. The normalized conductivity, $\sigma_0$, is
introduced for the case of short-range scattering, when $l_c = 0$.
The distribution $f_{p=0t}$ is shown in Fig. 4b for the case ($ii$)
while $2f_{p=0t}=1$ for the case ($i$). If $l_c=0$, one obtains
$\sigma_t/\sigma_0=1$, i.e. there is no transient photoconductivity
for the case ($i$); for the case ($ii$) one obtains
$\sigma_t/\sigma_0=2f_t$ and the transient photoconductivity is
clear from Fig. 4b.

\begin{figure}[ht]
\begin{center}
\includegraphics{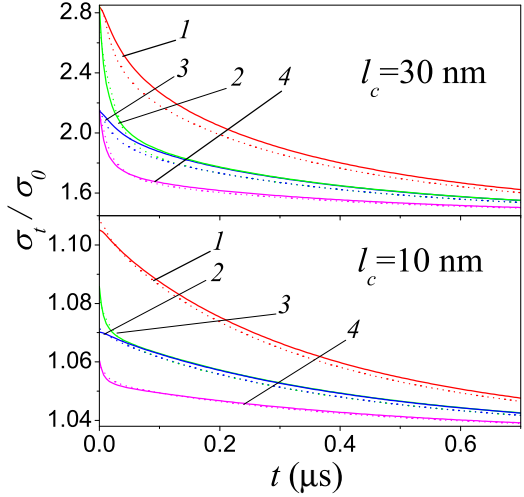}
\end{center}
\addvspace{-1 cm}
\caption{Temporal evolution of conductivity for excitation conditions (1)-(4) which
are the same as in Fig. 4 for the correlation length $l_c=$30 nm and 10 nm. Solid
and dashed curves are correspondent to the cases ($ii$) and ($i$). }
\end{figure}

If $l_c\neq 0$, the transient evolution of conductivity is shown in
Fig. 8. For the definiteness, it was assumed that $l_c$=10 and 30~nm
and variations of $\sigma_t$ are increased with $l_c$ essentially
due to contribution of high-energy carriers. Similar to Sec. IVA,
one can separate two stages of evolution: the fast decrease of
$\sigma_t$ due to energy relaxation (up to $\sim 30\div 50$ ns for
the conditions considered) and the slow quenching of $\sigma_t$ due
to carrier recombination. If $t>1~\mu$s, the conductivity approaches
to the equilibrium values: $\sigma_{t\to\infty} /\sigma_0=$1.445  if
$l_c$=30~nm and $\sigma_{t\to\infty}/\sigma_0=$1.035 if $l_c$=10~nm.
Since the transient conductivity can be measured for the
subnanosecond time scale \cite{21}, such a scheme can be used
for verification both energy relaxation and recombination
mechanisms.

\section{Concluding remarks}
To summarize, we have considered both the interband ultrafast
photoexcitation and the relaxation dynamics of the carriers in an
intrinsic graphene. In contrast to the measurements \cite{3,4,5,6}
and calculations \cite{8,9} performed, where the evolution
corresponds to the subpicosecond time scales due to the opticlal
phonon contribution, here we consider the slow relaxation of the
low-energy carriers. The distribution of carriers at $T=$77 K is
obtained for the limiting cases with negligible or dominating intercarrier
scattering when the energy relaxation and generation-recombination
processes are caused by the quasielastic acoustic phonon scattering
and thermal radiation, respectively. The initial distribution is
obtained in the framework of the linear, with respect to pumping,
approximation for the collisionless regime of the interband
transitions. The transient optical response on the probe radiation
(transmission and reflection) as well as on the weak dc field
(transient photoconductivity) appears to be strongly dependent on
the relaxation and recombination dynamics of carriers.

Next, we discuss the assumptions made. The main restrictions of the
results presented are the consideration of the low-energy carriers,
when the interaction with optical phonons is unessential, and the
single generation-recombination mechanism (due to thermal radiation)
is taken into account. These conditions are realized at low
temperatures under the mid-IR ultrafast excitation \cite{17} of the
clean sample (e.g. suspended graphene \cite{22}). Such an approach
can be used for the case of optical interband excitation, when the
low-energy initial distrbution, with a phenomenological broadening, is
formed after the cascade process of optical phonon emission. The
consideration is restricted by the radiative recombination (the
Auger processes are forbidden due to the symmetry of electron-hole
states \cite{23}), with the characteristic time scales up to
microseconds. Any visible contribution of other generation-recombination
mechanism (e.g., because of disorder-induced interband transitions
with acoustic phonons, or under intercarrier scattering) leads to fast
decrease of photoresponse. Such a regime requires an additional investigation
but the quasielastic energy relaxation stage is described by the
presented results.

The rest of assumptions are rather standard. The consideration in
Sec. III is limited by the simple cases ($i$) and ($ii$), with and
without the intercarrier scattering. The main peculiarities of the
response under consideration are similar for both cases but the
complete description of the nonequilibrium carriers had been
performed neither under optical excitation, nor under high dc field,
see \cite{18, 24} and Refs. therein. The description of the
momentum relaxation in Sec. IV is based on the phenomenological
model of Ref. 13. The utilization of the quasielastic energy scattering
and the collisionless interband photoexcitation appear to be rather
natural. The listed assumptions do not change either the character
of the response or the numerical estimates.

In closing, the peculiarities of the transient optical response
(transmission and reflection) as well as of the transient
photoconductivity appear to be useful tool in order to verify the
relaxation and generation-recombination mechanisms of carriers.
Thus, in addition to the recently obtained experimental results
\cite{3,4,5,6,7} measurements under mid-IR excitation and at
low-temperature will be useful for characterization of graphene.

\appendix
\section{Generation rate}
Below we describe the interband carrier excitation under ultrafast mid-IR
pumping ${\bf E}_t\exp (-i\Omega t)$ for the collisionless case, when $\tau_p$ is
shorter than relaxation times. The photogeneration rate into the $\alpha$-state
is based on the general expression (see \cite{1} and Sec. 54 in Ref. 12)
%A1
\begin{eqnarray}
G_{\alpha t}=2Re\left(\frac{e}{\hbar\Omega}\right)^2\int\limits_{-\infty}^0
d\tau e^{\lambda\tau -i\Omega\tau} ~~~~~~~~~ \\
\times\left\langle\alpha\left|\left[ e^{i\hat h\tau /\hbar}\left[\left({\bf E}_{t+\tau}
\cdot{\bf \hat v}\right) ,\hat\rho_{t+\tau}\right] e^{-i\hat h\tau /\hbar} ,
\left({\bf E}_t\cdot{\bf \hat v}\right)^+\right]\right|\alpha \right\rangle ,
\nonumber
\end{eqnarray}
where $\hat\rho_t$ is the density matrix, $\hat{\bf v}$ is the velocity
operator, and $\lambda\to +0$. Since the collisionless regime of photoexcitation,
we calculate (A1) with the use of the free states $|l{\bf p}\rangle$ and the
energy $\varepsilon_{lp}$ where $l=\pm 1$ stands for $c$- or $v$-bands and
$\bf p$ is the 2D momentum. Neglecting the nondiagonal components of the
density matrix $\hat\rho_t$ and using the distribution functions $f_{l{\bf p}t}$,
one obtains the generation rate
%A2
\begin{eqnarray}
G\{ f|1{\bf p}t\} =\left(\frac{e}{\hbar\Omega}\right)^2 \int\limits_{-\infty}^0
d\tau e^{\lambda\tau -i\Omega\tau}e^{i(\varepsilon_{1p}-
\varepsilon_{-1p})\tau /\hbar}  \nonumber  \\
\times\langle 1{\bf p}|({\bf E}_{t+\tau}\cdot{\bf \hat v})|-1{\bf p}\rangle
\langle -1{\bf p}|({\bf E}_{t}\cdot{\bf \hat v})^+|1{\bf p}\rangle ~~~~~ \\
\times \left( f_{-1{\bf p}t+\tau}-f_{1{\bf p}t+\tau} \right) + c.c. ~ , \nonumber
\end{eqnarray}
moreover $G\{ f|-1{\bf p}t\}=-G\{ f|1{\bf p}t\}$ according to the
particle concervation law. Next, we separate the envelope form-factor
$w_t$ using ${\bf E}_t={\bf E}w_t$ and take into account
the in-plane isotropy of the problem, when one arrives to the
averaged matrix element $\overline{\left|\left\langle +1{\bf
p}\left| ({\bf E}\cdot{\bf\hat v})\right| -1{\bf
p}\right\rangle\right|^2} =(Ev_W )^2/2$. As a result, we obtain the
in-plane isotropic generation rate $G\{ f|pt\} =\pm G\{ f|\pm lpt\}$
in the following form:
%A3
\begin{eqnarray}
G\{ f|pt\}=\left(\frac{eEv_W}{\hbar\Omega}\right)^2\frac{w_t}{2}
\int\limits_{-\infty}^0 d\tau w_{t+\tau}e^{\lambda\tau -i\Omega\tau}  \nonumber \\
\times e^{i(2v_W p)\tau /\hbar}\left( f_{-1{\bf p}t+\tau}-f_{1{\bf p}t+\tau}
\right) + c.c. ~ .
\end{eqnarray}
Finally, using the electron-hole representation and replacing the
filling factor here by $(1-2f_{pt})$, we arrive to Eq. (2).

\section{Dynamic conductivity}
The response of graphene on the in-plane probe field ${\bf E}\exp
(-i\omega t)$ is described by the dynamic conductivity \cite{20,25}
%B1
\begin{eqnarray}
\sigma_{\omega t}\approx i\frac{2(ev_W )^2}{\omega L^2}\sum\limits_{\bf p}
(1 - 2f_{pt} ) ~~~~  \\
\times\left(\frac{1}{\hbar\omega +2v_Wp+i\lambda}-\frac{1}{\hbar\omega
-2v_Wp + i\lambda} \right) \nonumber
\end{eqnarray}
with $\lambda\to +0$. The parametric time dependency of $\sigma
_{\omega t}$ is valid if the time scales under consideration exceed
$\omega^{-1}$. It is convenient to separate the time-independent
contribution, $\overline\sigma_\omega$, described the undoped
graphene in the absence of photoexcitation, when $f_{pt}$ vanishes.
Using the energy conservation law one obtains ${\rm Re}\overline
\sigma_\omega =e^2/4\hbar$. In the framework of the Weyl-Wallace
model, the ${\rm Im}$-contribution into $\overline\sigma_\omega$
appears to be divergent at $p\to\infty$. It is convenient to
approximate ${\rm Im}\overline\sigma_\omega$ as a sum
of $\propto\omega^{-1}$ and $\propto\omega$ terms, which correspond
to the contributions of the virtual interband transitions and the
ion background, correspondingly. As a result, we obtain:
%B2
\begin{equation}
{\rm Im}\overline\sigma_\omega\approx\frac{e^2}{\hbar}\left(\frac{\varepsilon_m}
{\hbar\omega}-\frac{\hbar\omega}{\varepsilon_i}\right) ,
\end{equation}
where the characteristic energies, $\varepsilon_m \simeq$0.1 eV,
and $\varepsilon_i \simeq$ 6.8 eV are correspondent to the recent
measurements of the graphene optical spectrum. \cite{26}

Next, substituting the time-dependent distribution $f_{pt}$ obtained
in Sec. III into the dynamic conductivity (B1) one transforms the real
and imagional parts of $\sigma_{\omega t}$ as follows
%B3
\begin{eqnarray}
{\rm Re}\sigma_{\omega t}=\frac{e^2}{4\hbar}\left[ 1-2F\left( p_{\omega},t
\right)\right] , ~~~~~~~  \\
{\rm Im}\sigma_{\omega t}={\rm Im}\overline\sigma_\omega -\frac{e^2}{\pi\hbar}
{\cal P}\int\limits_0^\infty \frac{dyy^2}{1-y^2}F( p_{\omega}y,t) . \nonumber
\end{eqnarray}
Here $\cal P$ means the principal value of integral. We also introduced the
function $F(p,t)=f_{pt}$ for the case ($i$) and
%B4
\begin{equation}
F\left( p_{\omega}y,t \right) =\frac{f_t}{\exp [(\hbar\omega /T_t)y] (1-f_t)+f_t}
\end{equation}
for the case ($ii$), when $\sigma_{\omega t}$ is determined both the effective
temperature and the carrier concentration, $T_t$ and $f_t$.


\begin{thebibliography}{52}
\bibitem{14}
The form-factor $w_t$ is normalized according to the condition $\int_{-\infty}^{\infty}
dtw_t^2=\tau_p$. The shape of normalized form-factor has little effect on the
transient photoexcitation under consideration because the ultrafast response
is determined fundamentally by the pulse duration, $\tau_p$.
\bibitem{15}
Here we use the characteristic velocities $v_{ac}\simeq$2.5$\times 10^5$ cm/s
(for the nitrogen temperature) and $v_r\simeq$41.6 cm/s (for graphene sheet placed
between SiO$_2$ substrate and cover layer), see \cite{10}.
\bibitem{19}
According to Eq.(B3), the negative interband absorption ($\propto{\rm Re}
\sigma_{\omega t}$) takes place, if $f_{pt}>1/2$, see \cite{10} and
Refs. therein. This low-energy instability is suppressed due to an effective
intraband (Drude) absorption.
\end{thebibliography}
\end{document}